\documentstyle{mn}
\begin{document}
 
\title{1.25 mm Observations of a Complete Sample \\
of IRAS Galaxies:
(II) Dust Properties ($\dagger$)}
\footnote {($\dagger$)} 
{\it Based on observations collected at the SEST 15m telescope (Chile).}

\author[Paola Andreani and Alberto Franceschini]{Paola Andreani$^1$ and 
Alberto Franceschini$^2$ \\
$^1$ Dipartimento di Astronomia di Padova, vicolo dell'Osservatorio 5,
 I-35142 Padova, Italy.
 e-mail: andreani@pdmida.pd.astro.it \\
     and  European Southern Observatory, D-85748 Garching, Germany\\
$^2$ Dipartimento di Astronomia di Padova, vicolo dell'Osservatorio 5,
 I-35142 Padova, Italy.
         e-mail: franceschini@astrpd.pd.astro.it}
 
\date{Accepted .....
      Received ..... in original form January 15th 1996}
%\pubyear{1996}
 
\maketitle
 
\begin{abstract} 
 
We present 1.25 $mm$ continuum data for a southern galaxy sample
selected from the IRAS PSC and complete to $S_{60}=2\ Jy$. Two thirds
of the galaxies have been detected and significant limits on the remaining
objects have been set. 
We find, on a statistical basis, indications that the dust emission
in these galaxies is somewhat more
centrally concentrated than that of the optical light, possibly tracing a
higher metal content in the inner galactic regions.
This result also allows to estimate the aperture corrections to
the millimetric data.
The latter, together with IRAS photometric data, have
been used to compare the broad-band FIR/mm spectra with a simple dust
model. According to their far-IR/mm spectrum, the sample galaxies show a
dichothomy: almost half of the objects, those displaying
bright 25-60 $\mu m$ fluxes
ascribed to warm dust residing in {\it starburst} regions, are
characterized by higher values of the bolometric (optical + FIR)
luminosity, of the dust-to-gas mass
ratio, of the dust optical depths and of the overall extinction.
A complementary class of objects dominated by cold dust ({\it cirrus}) 
shows opposite trends.
Because of the favourable
observational setup, selection wavelength and completeness, we believe these
data provide an exhaustive and unbiased view of dust properties
in spiral galaxies. 
 
\end{abstract}
 
\begin{keywords}
{ galaxies: ISM - galaxies: photometry - galaxies:
spirals, starburst - ISM: dust, extinction - infrared: ISM: continuum}
\end{keywords}

%\clearpage
%\newpage
%\vskip 2 truecm
 
\section{INTRODUCTION}
 
\noindent
Observations of diffuse dust in galaxies provide important information
on some basic questions about their present structure and past history.
Even minor amounts of absorbing dust play a crucial role in shaping the
galaxy's broad-band e.m. emission.
Any inferences about structural properties, such as the global baryonic 
content,
its spatial distribution, and the presence of a dynamically dominant
non-baryonic component, rely on our ability to account for the
missing fraction of light in optical searches (Disney, Davies and Phillips,
1989; Choloniewski 1991; Burstein, Haynes \& Faber
1991; Huizinga \& van Albada 1992; Giovannelli et al. 1994; Valentijn 1994;
Peletier et al. 1995). Are we missing
significant amounts of luminous matter because of the effects of dust 
extinction? How much severe are the corresponding selection effects? 
 
Dust is also an
important tracer of the activity of stellar populations in galaxies.
By cumulating the products of all stellar generations, diffuse dust provides
an integrated view of the past history of the object.
 
There is a tendency, as of today, to consider the innermost parts
of late-type galaxy discs close to optical thickness, but for
the outer regions the question cannot yet be settled with present data.
Fundamental questions of extragalactic astronomy,
e.g. the calibration of distance indicators, are affected by this lack 
of knowledge.
 
\noindent
Traditional approaches to investigate dust in galaxies rely on either 
{\it (a)} indirect estimates based on dust extinction effects in the optical 
and near-IR (e.g. White \& Keel 1992; James \& Puxley 1993; Block et al. 1994),
or {\it (b)} direct measurements of dust emission from FIR/mm observations.
Neither have provided conclusive results at the moment.
The former method is based on rather model-dependent and controversial 
assumptions about
the broad-band optical-IR spectrum of various galactic components. The
latter method, on the other hand, still lacks a large enough 
statistical basis and suffers from the observational uncertainties 
of the mm data.
 
IRAS survey data,
although having provided the starting point for studies of galaxy emission
at long wavelengths, are not enough by themselves 
to fully characterize the dust content and emission properties,
because of its limited sensitivity, spatial resolution, and spectral 
coverage. In particular,
dust colder than 20 K at large radial distances, whose existence is currently
matter of so much controversial discussions (Rowan-Robinson, 1992;
Valentijn, 1994; Block et al., 1994), can only be sampled at
$\lambda>100 \mu m$ (Chini et al., 1986; Thronson et al., 1987; Stark et al.,
1989; Andreani and Franceschini, 1992), where
direct observations are still limited to a handful of objects.
 
This paper reports on a millimetric continuum
survey of a complete sample of IRAS galaxies
performed at 1.25 {\it mm} with the 15m Submillimeter ESO-Swedish telescope
(SEST). A preliminary report on these observations has been given by
Franceschini and Andreani (1995).
 
\noindent
The SEST telescope
was chosen as it provided the best compromise between detector
sensitivity and spatial resolution. The 24$^{\prime \prime}$
FWHM of the SEST beam at 1.25 {\it mm}
and the average optical extent of the sample galaxies match favourably, so
that the beam-aperture corrections needed to compare with the IRAS survey
data are not as severe as for other observational setups.
 
\noindent
Observations at such long wavelengths, made possible by the high
sensitivity of the used bolometer system, ensure that all possible
dust components with
temperatures down to the fundamental limit of $T\simeq 3\ K$ set by the
cosmic background radiation are
properly sampled. The reference to a complete flux-limited sample ensures
a minimal exposure to the effects of bias, hence allowing any kinds of
statistical tests to be applied (determinations of IR-mm bivariate
luminosity distributions, luminosity functions and 
volume emissivities: Franceschini and Andreani, 1996, work in progress). 
Finally, the adoption of a far-IR selected sample,
rather than of an optical one, makes us confident that the whole
phenomenology of dust effects in galaxies is properly sampled.
 
Section 2 is devoted to the present observations and discusses the 
issue of the extent of the millimetric emission and the beam-aperture 
corrections. Section 3 reports results from the best-fit to the
observed spectra with a dust model and the inferred dust properties
for our sample, while the discussion is deferred to \S 4.
 
\section{OBSERVATIONS AND DATA ANALYSIS}
 
\subsection{The sample}
 
The galaxy sample under investigation 
was selected from the IRAS Point Source
Catalogue and is complete to the flux limit of
S$_{60 \mu m}=2$ Jy.
It consists of 30 galaxies with morphological types
from S0/a through Scd, within the sky region $21^h < \alpha (1950)
< 5^h$ and $-22.5^\circ <\delta<-26.5^\circ$. For most of these galaxies 
PSC data also at 100, 25 and 12 $\mu m$ were available.
Distances (mostly from distance indicators, otherwise derived using 
redshift data for $H_0=75$) cover the range from $d\simeq 20$ to $\simeq
200$ Mpc. 
Half-optical-light diameters $A_e$ are typically within $A_e\simeq
10^{\prime \prime}$
to $\simeq 40^{\prime \prime}$, four exceptionally close
 galaxies having $A_e \sim 100^{\prime \prime}$. 
The volume test ($<V/V_{max}>=0.45\pm 0.06$) ensures 
the sample completeness.
 
Table 1 reports some relevant information 
(name, coordinates, distances, optical dimensions,
blue magnitudes and morphological types) for the sample galaxies.
The data have been taken from the Lyon-Meudon Extragalactic Database 
(LEDA), the NASA Extragalactic Database (NED) and the
IRAS PSC (Moshir et al., 1989). Optical photometric data have been taken
from the ESO-LV catalogue (Lauberts \& Valentijn, 1989).

\vskip 1truecm
\centerline {\bf Table 1}
\vskip 1truecm
 
\subsection{The bolometer observations}
 
The sample has been observed at 230 GHz during various campaigns from
1990 to 1993 with the
SEST 15m telescope feeding a $^3$He-cooled bolometer of the MPIfR 
(Kreysa 1990). The diffraction-limited beam size is $24^{\prime
\prime}$ (FWHM).
Beam-switching is achieved by chopping
ON-OFF the source and nodding the telescope, which results in a three-beam
modulation on and off the source. The beam separation was set at
$70^{\prime \prime}$.
The pointing accuracy was most of the time $2^{\prime \prime}$ and was checked
each half a hour, by pointing a nearby radio-loud quasar.
Each source was observed $n \times 200$ seconds, with $n$ depending on the
expected 1.25 {\it mm} intensity (approximately evaluated by
extrapolating the IRAS $100 \mu m$ flux using a modified thermal spectrum
with spectral indices of dust opacity between 1 and 2).
 
The atmospheric 
transmission was monitored by frequent skydips. Uranus, Mars and Neptune
were used as primary calibrators and the pointing quasars
as secondary ones.
The overall accuracy of the measured fluxes (which we estimate to be
between 10 and 30\%) strongly depends on the atmospheric conditions. 
A more detailed description of the observing procedure 
can be found in Andreani \& Franceschini (1992) and Andreani (1994).
 
Table 2 reports our observed 1.25 {\it mm} fluxes (cols. 4 and 5),
together with the IRAS 
at 25, 60 and 100 $\mu m$ fluxes (cols. 1, 2, 3), 
all corrected for color and K-correction following Smith et al. (1987). 
In the lack of a detection a 3 $\sigma$ upper limit to the flux is reported.
 
\vskip 1truecm
\centerline {\bf Table 2}
\vskip 1truecm
 
\subsection{Beam-aperture corrections: the size of galaxies at
$\lambda=1.25\ mm$}
 
Corrections to the observed millimetric fluxes, to account for the beam
aperture of the telescope being comparable to or even smaller than
the angular size of the sources, are crucial to estimate the global
mm emission and to compare it with the
total fluxes provided by the IRAS survey. Such a difficult evaluation
is one of the main reasons
preventing so far to reach reliable conclusions about dust emission in
galaxies. 
 
Up to now our knowledge on the dust distribution in galaxies mostly 
relied on IRAS
data, whose basic results are briefly summarized in the following.
 
\noindent
Light profiles similar, on average, in the far-IR and optical band 
were inferred from a
direct comparison of blue-band, $\lambda=50$ and 100
$\mu m$ data using the IRAS CPC (Chopped Photometric Channel)
instrument by Wainscoat {\it et al.} (1987) on three nearby edge-on spirals
and by van Driel et al. (1995) on 55 nearby spirals. A study of
a sample of extended galaxies, partially resolved by IRAS (Rice
{\it et al.} 1988), shows that the mean ratio of the far-IR (60 $\mu
m$) $D_{IR}$ to
blue-light $D_B$ isophotal diameters turns out to be 0.98$\pm$0.25.
This implies that on average galaxies have far-IR extensions comparable to
the blue ones. A similar result is found by
Xu \& Helou (1995) for M31. They combined FIR and UV data of M31 and
a dust heating/cooling model, to study the extinction in the disc of this
galaxy. The overall distribution of V-band
optical depth in this galaxy is rather flat and it is very tight 
correlated with the HI column density.
There are IRAS objects, however, where the IR emission seems to be
more centrally concentrated than that of the blue light. 
 
First attempts to map mm dust emission have been recently done by Chini \&
Kr\" ugel (1993) and Chini et al. (1995), who have shown that the
distribution of the millimetric
light is similar to the optical one and argue that for the observed objects
cold dust emission is extended as far as the blue light.
 
Two main advantages distinguish our sample:
the majority of the objects are relatively distant and have small
angular sizes in the optical. Furthermore, the SEST antenna has a
a relative large beam-width with respect to other millimetric telescopes
(JCMT and IRAM). 
In spite of this, aperture corrections may still turn out to be significant
and deserve particular attention. 
The solution we devised is the following.
For some of the optically most extended sources 
(0300--23, 0335--24, 2217--24, 2233--26) we have performed
a rough mapping by integrating on four points, one beam-width distant,
around the galaxy center. The fluxes listed in Table 2 have been
obtained by adding the contributions of each
observed position. This allows at least to 
reduce the large aperture corrections for these sources.
 
\noindent
The galaxy 0045-25 (NGC 253) was not observed by us and the flux
listed in Table 2 is taken from Chini et al. (1984).
These authors performed aperture photometry with a beam of width of
3.9$^\prime$ and measured a flux of $4.4\pm1.7$ Jy.
The inner region (2$^\prime$x3$^\prime$) of this object was
also mapped at 1.25 {\it mm} by Kr\"ugel et al. (1990). These authors
estimate a total flux in this inner region of 1.7$\pm$0.2 Jy. However, this
well-known object is very nearby and therefore quite extended in the optical
(27.4$^\prime$x6.8$^\prime$), it is very likely that both measurements
have underestimated the real emission.
 
More generally, our adopted procedure was to test three 
different hypotheses about the geometrical distribution of the millimetric flux.
(1) Sources are point-like objects, hence no beam-aperture
correction is to be applied; (2)  
the radial distribution of the millimetric light is exponential with a
scale-length $\alpha_{\rm mm}$ equal to one third of the optical one:
$\alpha_{\rm mm} = \alpha_0 / 3$ (case of dust emission 
more concentrated than the star-light) and (3) the dust emission follows
that of the optical light ($\alpha_{\rm mm}= \alpha_0$).

\noindent
The model surface brightness distributions corresponding to the three
cases have been convolved with the telescope beam shape following 
Andreani and Franceschini (1992).
Results of these aperture corrections to the observed mm fluxes are reported 
in Table 2. In columns
6 and 7 the fluxes corrected according to hypothesis (2) are listed,
and in columns 8 and 9 those corrected according to hypothesis (3).
Note that for case (2) the aperture-correction
turns out to be 20\% on average of the observed mm flux 
(the largest correction is a factor 3 for 0311-25). An average 
correction of roughly a factor 2 is instead required according to 
hypothesis (3).

\noindent
We have tested further the three different hypotheses on the {\it mm} light
distribution against our data as follows.
Figures 1a and 1b are plots of the ratios of the 1.25 {\it mm} to the 60 and 
100 $\mu m$ fluxes 
versus galaxy distance and diameter, with $S_{1.25}$ corrected
according to the three above mentioned hypotheses.
The two lines in each panel of Fig. 1 correspond to the $\pm 1\ \sigma$
regression fits obtained with a "survival analysis" technique.
 
\begin{figure}
 \vspace{302pt}
\caption{(a) Millimetric to far-IR flux ratios versus galaxy distance.
Top panels
compare with the 100 $\mu m$ flux, bottom panels with the 60 $\mu m$.
In panels to the left the mm data have been corrected for beam-aperture 
assuming point-sources, central panels assume a mm scale-length of 1/3 of the
optical one, right-hand side panels assume the same distribution for the mm 
and optical light. Panel (b) same as (a), but for the far-IR flux ratios
versus galaxy diameter.}
\end{figure}

We see in Fig. 1a that hypothesis (3) in particular implies 
a dependence of the average flux 
ratio with distance which is significant at the 3.5$\sigma$ level for 
the 100 $\mu m$ (and 5.5$\sigma$ for the 60 $\mu m$ match).
This is of course  not expected,
since all flux measurements have already been K-corrected. 
We are led to conclude that the hypothesis of the presence of {\it emitting}
dust at large radial distances appear to be inconsistent with our data,
as detailed in Sect. 3 below.
Instead, the case
for a higher concentration of dust emission, with $\alpha_{\rm mm}$ between
$\alpha_o/3$ and $\alpha_o/2$, is fully supported.
An even stronger indication in favour of the latter assumption 
is observed in Fig. 1b, where the mm-IR flux ratios versus
the galaxy effective diameter is shown. 
 
\noindent
The millimeter fluxes used hereafter
were therefore corrected for the effect of beam-aperture according 
to the assumption $\alpha_{\rm mm} = \alpha_0/3$.
Our mm data for all the sample galaxies, as well as the complementary IRAS
fluxes, are plotted in Figure 2. Best-fits of a dust emission model
discussed in the next Section also appear in the figure.

\begin{figure}                  %#2
 \vspace{602pt}
\caption{FIR/mm spectra of the sample objects. In each panel the name, the
total FIR/mm luminosity and dust mass (estimated from the dust model) are
shown. Solid lines are the global fitted spectrum, including two dust
components: (1) the {\it cirrus} (dashed line), and the 
{\it starburst} component (dot-dashed line).} 
\end{figure}
 
\section{THE DUST CONTENT OF GALAXIES}
 
\subsection{The dust model}
 
Data at millimeter wavelengths, improving on the spectral dynamic range 
with respect to the IRAS
survey data, add significantly to our knowledge of dust 
properties in infrared galaxies. No dust component, at whichever
temperature, is expected to escape a combined mm-IRAS survey. 
 
FIR spectra of spiral galaxies are schematically interpreted as
due to thermal  re-radiation from dust present in three main 
galactic environments: {\it (a)}
dust grains heated by the general interstellar radiation field ("cirrus"
emission including cold dust and small transiently-heated
hot dust grains), {\it (b)} dust heated
by hot stars in extended HII regions, and {\it (c)} dust in 
compact HII regions
and molecular clouds. 
The "cirrus" component has been modeled following the recipes 
by Rowan-Robinson (1986, 1992; hereafter RR86, RR92). 
The model predicts the existence 
of nine grain types: amorphous carbon and silicate grains of size
$a=0.1\ \mu m$, graphite and silicate of 
0.01 and 0.03$\mu m$ size, very small 
($a=0.002$ and 0.0005 $\mu m$) transiently-heated graphite grains, 
and finally
a component of very large amorphous grains with radius $a=30\ \mu m$ to 
account for the observed submillimeter properties. The mixture and
grain properties have been optimized by RR92 to reproduce all basic 
observables (extinction law, absorption properties of the ISM towards 
various 
galactic regions) over the entire 1 mm to 0.1 $\mu m$ spectral range.
 
For normal grains the emission temperature is assumed to attain equilibrium
(the thermal capacity is large enough with respect to the energy of the field
photons), and is estimated from an energy balance equation between the emitted
and absorbed photons (including the scattered fraction).
On the contrary, very small grains, because of their small sizes,
are not in thermal equilibrium with the impinging radiation
field and their emission fluctuates according to their changes in temperature.
Their thermal behaviour is described by a probability distribution
function P(T),
which has the approximated analytic expression given by Draine
\& Anderson (1985):
 
$$ P(T) = \cases {6.680\cdot T^{-2.75} &if $2.7\leq T \leq 80 ~~~ a =20 \AA$;
\cr 0.168\cdot T^{-2.75} &if $2.7\leq T \leq 500 ~~~ a =5 \AA$} $$
 
\noindent
Their scattering and absorbing properties
for $\lambda > 0.1 \mu m$ are those of 0.01$\mu m$ grains scaled by
a factor of $a$/0.01 (see below). In fact for small grains ($a\leq 0.1\lambda$)
the extinction efficiency, $Q_{ext}/a$ is independent of $a$.
 
For the warm dust in star-forming regions we have followed a simplified but
successful approach as illustrated by Xu and De Zotti (1989) and Conte (1993).
The star-forming region is modeled as a volume uniformly filled with
hot stars, dust and radiation, and is assumed to be optically thin at least
for $\lambda > 20\ \mu m$. The same grain mixture as for the "cirrus" 
component is adopted, but very small grains are assumed to be destroyed
by the intense radiation field.
 
\vskip 0.2truecm
\subsection{The dust emissivity}
 
The total emissivity due to the hot, warm and cold components is given
respectively by:
 
$$ 4\pi \epsilon _T (\nu) = $$
$$ \sum _{i=1}^2 \chi_c \int 4\pi N_{gr,H}(a_i) Q_{ext}(a_i,\nu)
\pi a_i^2 P(T) B_\nu(T)dT +$$ 
$$ \sum_{i=1}^4 4\pi N_{gr,H}(a_i) Q_{ext}(a_i,\nu) \pi a_i^2 B_\nu(T_i) +
\eqno(1)$$
$$ \sum_{i=1}^7 4\pi N_{gr,H}(a_i) Q_{ext}(a_i,\nu) \pi a_i^2 B_\nu(T_i)
$$
 
\medskip\noindent
$N_{gr,H}$ are the number of grains per H atom, $Q_{ext}(a_i,\nu)$ are
the extinction efficiencies, $B_\nu(T_i)$ is the black body spectrum at
the equilibrium temperature and
$\chi _c$ is the ratio between the intensity of the interstellar radiation
field and that of the field in the solar neighbourhood.
This latter has been taken by Mathis et al. (1983).
The sum of the warm component runs over only four of the seven components,
because for warm environments, such as those of HII regions,
the differences in optical properties of grains
with equal dimensions and different chemical compositions can be neglected.
In this case, the averaged extinction efficiencies and 
total abundances have been put in eq.(1). 
The UV absorbing efficiency for these grains
was taken equal to 1 and the radiation field was assumed uniform
inside the HII regions.

The equilibrium temperatures for all grains of both the "cirrus" and starburst
components are uniquely determined by solving the balance equations with the
the average radiation field intensity, $\chi_c$, in the cold "cirrus" and
the intensity illuminating the warm dust in star-forming regions, $\chi_w$,
as free parameters.
The third parameter used in the model is the light fraction $f_w$ at
100 $\mu m$ contributed by warm dust to the galaxy spectrum.
These three parameters fully describe the dust model.
The fitting procedure assumes that the 1.25 $mm$ flux
is dominated by cold dust belonging to the "cirrus" component, which mostly 
defines $\chi_c$. The 60 and 25 $\mu m$ data mostly define the other two
parameters.

\subsection{Dust distribution in galaxy discs}

We test in this section which spatial distributions of dust in galaxy
discs correspond to the surface brightness distribution of the millimetric 
emission inferred in \S 2 from our observations (though we recall that
this distribution has been indirectly estimated using a statistical
argument, and then may be prone to unaccounted systematic effects).
We expect, in general, that because of the scaling with radius of the
interstellar radiation field and hence of the dust emissivity, the $mm$
light should be more centrally concentrated than the optical.

\noindent
We have modelled the galactic interstellar radiation field as an
exponential scaling with the radial distance 
in the disc $I=I_0\cdot exp(-r/r_d)$
(with $r_d$=3.5 kpc, e.g. Reid 1994), i.e. closely following the distribution
of stars. We have then computed the dust emission according to our
adopted dust model, assuming {\it (a)} that the dust distribution
is more centrally concentrated (with scale-length of one third of
that of stars) and {\it (b)} alternatively that it follows the starlight
distribution. 

\noindent
The integrated surface brightness distributions
resulting from this calculation are shown in Figure 3, the solid curve
referring to the interstellar radiation field, the dashed one to a
$mm$ emission resulting from a dust distribution equal to that of
stars, the dotted curve corresponding to the case of a 
dust scale-length of one third of that of stars.
The corresponding ratios of the optical scale-length, $\alpha_o$,
to the {\it mm}-light scale-length, $\alpha_{mm}$, are 1.3 in the first
case and 3.1 for the second. Obviously, for a centrally 
concentrated dust distribution, the key factor is $\alpha_{mm}$ rather 
than $\alpha_o$. This implies that the spatial distribution of 
the {\it mm} light that we have inferred roughly coincides with the
real underlying dust distribution.

\begin{figure}
 \vspace{200pt}
\caption{Integrated galactic light as a function of the galactic radial distance.
The solid curve corresponds to the starlight, the dashed curve to the dust
emission found by assuming the dust distribution follows that of the stars.
The dotted curve represents the integrated dust emission with a
dust distribution 
more concentrated that that of stars (the scale-length in this case is one third
of the optical).}
\end{figure}
 
\subsection{Best-fit dust models to the observed spectra}

The best-fit spectra for the sample galaxies are shown in Fig. 2.
The adopted dust model, for a suitable choice of the parameters,
reasonably reproduces the
observed far-IR/mm broad-band spectra for all objects,
a result which confirms and extends previous findings by RR92.
The dashed line in Fig. 2 corresponds to the "cirrus" component,
the dot-dashed ones to the starburst
component. 
 
For 16 galaxies in our sample the contribution of warm 
dust at 100 $\mu m$ appears to be negligible, and sometimes not even 
required to improved the fit with a pure "cirrus" component. We call
these {\it "cirrus"-dominated} galaxies. 
On the other hand, 14 objects require high values of the warm dust 
fraction ($f_w\geq 0.3\div0.4$), and their spectrum at
$\lambda < 100\ \mu m$ is 
dominated by the starburst component (the {\it starburst-dominated} 
galaxies). We then suggest that 
far-IR to mm data and a reliable dust model
provide a straightforward way to classify galaxies according to their
star formation activity. This allows a complementary classification tool
with respect to the usual approach based on optical-UV
spectro-photometry.
 
In each panel of Fig. 2 best-fit values for the total dust mass 
and the FIR luminosities are also reported.
The latter have been computed by integrating the total FIR/mm spectrum.
Dust masses for both cold and warm dust have been estimated following
RR86's recipes for the mean density of grain material, 
while, for a given spectral fit, the total number of grains is set by 
the far-IR flux normalization. The mass of diffuse dust in the
interstellar medium is found to be mostly
contributed by the cold "cirrus" component (97.5\% on average), 
with only a minor fraction of the mass resident in the starburst 
component. The dust fraction involved in the latter ranges from typically
several percent for starburst galaxies to one percent in the inactive
population.
Figure 4 shows the average FIR/mm spectrum for 30 galaxies in
our sample fitted with a two-dust component and the corresponding best-fit 
parameters of the dust model are listed in Table 3 (the reduced
$\chi^2_\nu$ is 0.68). Note that
the starburst component contributes only a couple of percent
on average of the total dust mass.
 
\vskip 1truecm
\centerline {\bf Table 3}
\vskip 1truecm

The quoted dust masses should be taken as lower limits:
some dust may be hidden at low
enough temperatures, given the strong dependence of dust emissivity on
temperature. We argue here that only very sensitive {\it mm} observations
(like those presented here) can tackle it successfully. In fact,
though the results from these fits are by no means unique, there is no
room for large amounts of dust mass in the framework of the adopted
model. The data are 
compatible with models having an additional very cold dust component
illuminated by a very dim radiation
field ($\chi_c = 0.5 \chi _\odot$) and with a temperature distribution
ranging from T$\sim$4 to 18 K. In this case this very cold component would
bear 76.3 \% of the total mass and the total average dust mass would
be increased by 30\%.

\noindent
The largest amount of dust still allowed by our observed dust spectra
may be computed assuming that all
dust components are set to the fundamental temperature of 3 K (the
corresponding radiation field would be $\chi_c = 0.01 \chi _\odot$).
%In such a case the microwave background photons would be responsible
%for eating the dust grains.
The resulting mass of this very cold dust would include 90\% of the 
total, which would be twice larger than our best-fit value.
Then we estimate that our dust masses are correct within
a factor of two, for the adopted dust model.

\begin{figure}                    %#3
 \vspace{302pt}
\caption{The average far-IR millimeter broad-band spectrum for the
{\it median} galaxy in our sample.
%Two dust model fits are reported.
%In panel (a)
A two component model, including the
{\it cirrus} (dashed line) and {\it starburst}
(dot-dashed line) contributions with best-fit parameters $\chi_c =
11 \chi_\odot$, $\chi _w = 120 \chi _\odot$ (normalized to solar
neighborhood)
%and $f_w=0.3$, is shown. In panel
%(b) we report a three component fit with with the corresponding {\it cold
%cirrus} (dashed line) {\it very cold
%cirrus} (dotted line) and {\it starburst}
%(dot-dashed line) contributions. The best-fit parameters are $\chi_{vc} =
%0.5 \chi_\odot$ (radiation field intensity illuminating the very cold dust
%component), $\chi_c = 20 \chi_\odot$,
%$\chi _w = 120 \chi _\odot$.
The flux normalization at 100 $\mu m$
of the starburst component is $f_w=0.3$.
%that of the very cold dust is $f_{vc}=0.012$.
The physical parameters of dust grains are detailed in Table 3.
The median value of the 100 $\mu m$ flux for our sample galaxies is
$S_{100}=7.5\ Jy$.
}
\end{figure}
 
\subsection{Effects of dust extinction on the overall spectrum}
 
An overall view of dust effects in our IR
galaxy sample is summarized in Figure 5 in terms of extinction
versus dust optical-depth. The observed B-band optical depth $\tau_B$
has been estimated from the total dust mass divided by the projected area
within which we estimate to be resident the cold dust, i.e. the 
inner third of the optical radius (R$_{25}$), properly scaled with the
axial ratio squared $(b/a)^2$ to account for the disk inclination.
$\tau_B$, which measures the amount of dust {\it available} to absorb 
the optical light, is compared in Fig. 5 with the
quantity $A_B$ measuring the overall 
{\it actual}
effect of extinction. The average B extinction, $A_B$, has been 
estimated from the logarithmic ratio of the bolometric optical-UV 
luminosity from 0.1 to 10 $\mu m$ ($L_O$) to the bolometric far-IR light
($L_{FIR}$, 10 to 1000 $\mu m$) (two objects are missing in Figure 5
because optical photometry is not available).

\begin{figure}                %#4
 \vspace{202pt}
\caption{Observed optical depth of dust $\tau_B$ versus the overall 
extinction $A_B$. $\tau_B$ is averaged within one third of the 
R$_{25}$ radius. $A_B$ is estimated by comparing the bolometric outputs
in the optical and far-IR. Open and filled squares refer the the 
inactive {\it cirrus}-dominated and to the starbursting objects.
The position marked by the symbol ($\odot$) corresponds to our Galaxy.
}
\end{figure}
 
We see that the two galaxy classes, that we have defined are significantly
segregated over this plane, the inactive "cirrus"-dominated objects being
confined to lower values of dust optical depth ($\tau_B < 2$) and low
extinction ($A_B \leq 1$). The active star-forming galaxies, on the
contrary, are spread over much larger values in both axes. 
The predicted dependences for a screen, a slab, and a sandwich (with zero
scale-height of dust) model are shown for comparison (see Disney et al., 
1989).  Appreciable amounts of dust and  extinction seem
to characterize only a population of starbursting galaxies, with typical
values of the effective optical depth $\tau_B$ of 1 to 20 in the inner
one third of the optical radius. However, only for a minority of these 
starbursts
(4/13) extinction values significantly higher than 1 are indicated.
 
The solar symbol ($\odot$) in the plot of Fig. 5 marks the position 
of the Galaxy. We have estimated it by
applying the model of \S 3.1 and \S 3.2 to the Galactic
integrated spectral emissivity, as shown in Figure 6. The Galaxy spectrum is
well fitted by a pure cirrus emission with $\chi_c/\chi_{\odot} \simeq 7$.
The estimated total dust mass 
turns out to be $10^7$ M$_\odot$ while the optical depth 
$\tau \sim 0.3$. These values are somewhat lower than estimated by 
Sodroski et al. (1994), 
because of our different fit to the long-wavelength part of the spectrum 
(see comparison Fig. 6).
The position of the Galaxy within the region occupied by
the inactive galaxy population is consistent with our finding that the
Galaxy's far-IR spectrum is dominated by cold {\it cirrus} emission.

\begin{figure}
\vspace{202pt}
\caption{Best-fit of the dust model to spectral intensity data on our 
Galaxy (Puget 1989; Page et al. 1990; Wright et al. 1991). 
The data are fitted by a pure {\it cirrus} emission (no
starburst emission), with $\chi_c=8 \chi_\odot$. The total dust mass is
$M_d=10^7\ M_{\odot}$. The dashed line corresponds to the best-fit by 
Wright et al. (1991).
}
\end{figure}
\subsection{The dust-to-gas mass ratio}
 
We report in Figure 7 a plot of the gas (HI + molecular, from
Andreani et al., 1995) to dust
ratio versus bolometric luminosity ($L_{bol}=L_O + L_{FIR}$). We first notice 
that the median $L_{bol}\sim 4.5\ 10^{10}\ L_{\odot}$ for the inactive objects
is a factor 2.6 lower than the corresponding value for starbursts, consistent 
with our inference that the latter are characterized by an enhanced
star-formation activity, hence higher luminosities.
Second, the median gas/dust ratios for the two classes also differ on
average
($<M_g/M_d> \simeq 400$ for the starbursts and $\simeq 1000$ for the
inactive population). Finally, there is an apparent trend of $<M_g/M_d>$
to decrease with $L_{bol}$ from values of 3000 to less than 400 for the
inactive population.
 
The gas-to-dust ratio is particularly large ($\geq$3000) for three 
quite extended and close-by objects:
0300-23, 0045-25 (NGC 253) and 0128-22. The most likely explanation is that 
we have missed part of the dust emission, or underestimated the aperture
correction to be applied
(see \S 2.3), for 0300-23 and 0128-22, and it is probably even more
so for the observations of NGC 253 by Chini et al. (1984) and by
Kr\"ugel et al. (1990), whose photometry refers only to the inner region.
In principle, it could be argued that these three objects
are truly metal poor, but this is unlikely since
CO emission were detected in all of them (Andreani, Casoli \& Gerin
1995; Devereux \& Young, 1990).

\begin{figure}
\vspace{200pt}
\caption{The gas to dust mass ratio versus bolometric optical/mm luminosity
(symbols as in Fig. 5). Mass of gas (including both HI and molecular gas)
estimated from CO observations (see Andreani et al., 1995).
}
\end{figure}
 
The position of the Galaxy in Figure 7 has been found by scaling the value
of the ratio $<M_g/M_d>$ of Sodroski et al. (1994) 
to be consistent with our lower estimated $M_d$. 
It is confirmed that the Galaxy tends to have a somewhat low gas-dust ratio,
but the difference with respect to the median value of inactive
galaxies of the same luminosity is now a factor 2, only marginally
significant.

\section{DISCUSSION}

A fruitful combination of good sensitivity at long wavelengths, a large
beam aperture and small angular size of the target objects, have allowed 
us to explore the dust content in galaxies down to the coldest possible 
temperatures and over a significant fraction of the optical galaxy extent.

Although our inferences on the millimetric size of galaxies have only a
statistical sense, our results in Fig. 1 appear inconsistent with the
assumptions that cold dust emission has a scale-length comparable to or
larger than that of starlight. They rather suggest that the {\it mm}
scale-length is somewhat smaller than the optical one, though is not
possible to precisely quantify the difference until large bolometer
arrays will be available. 

\noindent
From a detailed modelling of dust emission from a galaxy in which the dust
is illuminated by an exponentially distributed radiation field, we
conclude that our inferred difference between optical and $mm$ lights
corresponds to a dust distribution which is more centrally concentrated
than the starlight, with a typical scale-length of one third of that of
stars. 

\noindent
This effect may bear some relationship with the observed 
increase in metallicity towards 
the inner regions of galaxies (e.g. Issa et al. 1990; Sodroski et al., 
1994; Carollo, Danziger, Buson, 1995) and with the idea of 
an enhanced past stellar activity there.
A similar effect is seen in the Galaxy and in M31, where a dust-to-gas ratio 
increasing towards the inner Galactic regions is observed
(Sodroski et al. 1995; Xu \& Helou 1995). This fact
is related by Sodroski et al. to the metallicity gradient
as a function of the galactocentric distance (Panagia \& Tosi, 1981;
Shaver et al. 1983; Wilson \& Rood 1994) observed in the Galaxy.
 
A published model (RR86, RR92) reproduces quite well the observed
broad-band spectra (see Figures 2 and 4). According to their FIR/mm spectrum, a
simple color criterion, supported by model predictions, allows us to 
classify these IR galaxies in two different groups: an {\it inactive} galaxy
population, the we called {\it cirrus} dominated objects, and an active
star-forming population, called {\it starbursts}.

\noindent 
The average galaxy FIR/mm spectrum can be fitted by a dust model
including contributions from a cold and warm dust components, 
with the cold one bearing most
of the dust mass. We find for the {\it median} galaxy in our sample
a value of $1.9\ 10^7\ M_\odot$ in dust, only 2 percent of which is warm
dust associated to star-forming regions. 

\noindent 
We have tested the presence of very cold dust, still in the frame of
the adopted dust model, by putting all dust components to the fundamental 
temperature of 3K (illuminating it with a very low radiation-field intensity).
This, probably extreme, condition would increase the average dust mass to
$3.5\ 10^7\ M_\odot$, 90\% of which in very cold dust. We argue here that
in the frame of the adopted dust model there is no large room for hidden cold
dust and the estimation of the dust mass is quite robust. 

It must be noted that estimation of the dust masses based on IRAS data 
{\it only} would produce values smaller by one order of magnitude.
This was already noted by other authors. For instance, even indirect methods
used to infer the amount of dust in galaxy disks, such as those based on
the observed B-K and V-K colours, together with population synthesis models
(Block et al. 1994), or the
combination of FIR and UV data with a dust heating/cooling model (Xu and 
Helou, 1995),
agree in finding dust amount larger than that inferred on the basis
of the IRAS data alone.
The dust mass evaluations for the two spirals NGC 4736 and NGC 4826
and for M31 are in agreement with the average dust
mass found in the present work, while the gas-to-dust ratio is $\sim$ 100,
as in the Galaxy.
 
The average B-band optical depth along the
line-of-sight, found by assuming that the total observed dust mass is
confined within one third of the optical radius,
spans a large range of values ($\tau_B=0.1$ to 50, see Fig. 5). 
The largest values of dust mass, optical depth ($\tau_B>1$) and of
the dust/gas
mass ratio are shared by objects characterized by a starbursting
activity.
The inactive "cirrus"-dominated galaxies, which are the typical population
that bright optical samples select, seem to be not much obscured by dust.
 
\noindent
These results are confirmed by an inspection of the {\it actual} $A_B$
extinction values
inferred from the observed ratio of optical to far-IR bolometric emissions.
Relatively high values of $A_B$ ($>1$) are displayed only by a minority of 
the starbursting galaxies (6 out of 28 objects). Note, however, that for
the latter class of sources the quantity $A_B$ is probably quite a 
poor measure of the real extinction affecting the optical galaxy, since
most of the far-IR emission there comes from starburst regions occupying
very small volumes in
the galactic body. In these cases too the average
extinction is probably not larger than 1 mag.
 
There are some intriguing effects to notice in Fig. 7. The starbursting 
activity seems to imply a heavier dust enrichment of the ISM with respect to
inactive spirals. The $M_g/M_d$ median values are $\simeq 1000$ for
the latter and 400 for the starbursts. The ongoing starburst
may well have preprocessed the ISM in these objects significantly longer 
or more efficiently than inactive galaxies did.
 
\noindent
The somewhat low $M_g/M_d$ value found for the Galaxy in Figure 7 with 
respect to other {\it cirrus} dominated objects in our sample 
may be explained by the fact that the
value of the ratio for the Galaxy mostly refers to the inner region, 
where most of the dust is concentrated, because such region is
more strongly weighted in the observations of the Galactic emission. 
On the contrary, observations of external 
galaxies include the contribution of the outer dust-poor environments.
So, we probably have to take into account a systematic bias when comparing
gas-to-dust ratio measurements for the Milky Way and for distant galaxies.
%However, if dust is more extended than assumed in the present work, the
%overall amount of dust would be higher by a factor of 2, on average, and
%thus the gas-to-dust ratio correspondly smaller and no significant
%differences would be detected with the Galactic value.
 
\noindent
Finally, there seems to be a trend favouring higher values of the mass in
dust $M_d$ relative to the total gas content $M_g$ in higher luminosity
{\it normal inactive} spirals. Assuming that this trend just reflects an
increased metallicity, this effect parallels a similar trend of higher
metallicities at higher masses inferred for galaxies of any morphological
types (see a recent review in Edmunds and Terlevich 1992).
 
\noindent
A number of abundance indicators, from the specific $Mg_2$
to the more general and widespread color-magnitude relationship, 
show that larger
fractions of supernovae debris are retained at greater galaxy masses
(probably due to the deeper gravitational wall). An effect of this kind
may be seen in our observations of the dust fraction in IR 
galaxies. The fact that starbursting galaxies in Fig. 7 do not follow the
same relationship may be due to a much different history of metal
production.

\section*{Acknowledgments}
Dave Clements is warmly thanked for taking the data of three objects.
We are also grateful to Phil James, the referee of this paper, for his
valuable comments and suggestions.

\end{document}